\documentclass[iop]{emulateapj}

\usepackage{natbib}
\usepackage{ulem}
\usepackage{comment}
\usepackage{txfonts}
\usepackage{color}

\usepackage[T1]{fontenc}

\usepackage{bm}

\newcommand{\eg}{e.g., }
\newcommand{\ie}{i.e., }
\newcommand{\Msun}{M_{\odot}}
\newcommand{\Rsun}{R_{\odot}}
\newcommand{\kms}{km~s$^{-1}$ }
\newcommand{\cms}{cm~s$^{-1}$ }
\newcommand{\ergs}{erg~s$^{-1}$ }
\newcommand{\years}{yr$^{-1}$ }

\newcommand{\Nifs}{$^{56}$Ni~}

\def\gsim{\mathrel{\rlap{\lower 4pt \hbox{\hskip 1pt $\sim$}}\raise 1pt
\hbox {$>$}}}
\def\lsim{\mathrel{\rlap{\lower 4pt \hbox{\hskip 1pt $\sim$}}\raise 1pt
\hbox {$<$}}}

\newcommand{\si}{{\sc i}}
\newcommand{\sii}{{\sc ii}}
\newcommand{\siii}{{\sc iii}}

\newcommand{\ej}{{\rm ej}}

\newcommand{\bkv}{{\rm 16bkv}}
\newcommand{\EoP}{{\rm EoP,sn}}
\newcommand{\Z}{{\rm 03Z}}
\newcommand{\csm}{{\rm CSM}}

\newcommand{\sone}{$s_{{\rm 1}}$}
\newcommand{\Ha}{H$\alpha$ }
\newcommand{\Hb}{H$\beta$ }
\newcommand{\hs}{\hspace{-1.5mm}}
\newcommand{\habs}{\hspace{-1mm}}
\newcommand{\bvri}{{\it BVRI}-band }
\newcommand{\jhk}{{\it JHK$_{s}$}-band }

\newcommand{\blue}{\textcolor{black}}

\shorttitle{Low-Luminosity Type IIP SN~2016bkv}
\shortauthors{Nakaoka T., et al.}

\begin{document}

\title{Low-luminosity Type IIP Supernova 2016bkv with early-phase circumstellar interaction}
\author{
  Tatsuya Nakaoka\altaffilmark{1},
  Koji S. Kawabata\altaffilmark{1,2},
  Keiichi Maeda\altaffilmark{3},
  Masaomi Tanaka\altaffilmark{4},
  Masayuki Yamanaka\altaffilmark{2,5},
  Takashi J. Moriya\altaffilmark{4},
  Nozomu Tominaga\altaffilmark{5},
  Tomoki Morokuma\altaffilmark{6},
  Katsutoshi Takaki\altaffilmark{1},
  Miho Kawabata\altaffilmark{1},
  Naoki Kawahara\altaffilmark{1},
  Ryosuke Itoh\altaffilmark{1,7},
  Kensei Shiki\altaffilmark{1},
  Hiroki Mori\altaffilmark{1},
  Jun Hirochi\altaffilmark{1},
  Taisei Abe\altaffilmark{1},
  Makoto Uemura\altaffilmark{1,2},
  Michitoshi Yoshida\altaffilmark{2,8},
  Hiroshi Akitaya\altaffilmark{2,9},
  Yuki Moritani\altaffilmark{2,10},
  Issei Ueno\altaffilmark{1},
  Takeshi Urano\altaffilmark{1},
  Mizuki Isogai\altaffilmark{4,11},
  Hidekazu Hanayama\altaffilmark{12}, and
  Takahiro Nagayama\altaffilmark{13,14},
}

\altaffiltext{1}{Department of Physical Science, Hiroshima University, Kagamiyama 1-3-1, Higashi-Hiroshima 739-8526, Japan; nakaoka@astro.hiroshima-u.ac.jp}
\altaffiltext{2}{Hiroshima Astrophysical Science Center, Hiroshima University, Higashi-Hiroshima, Hiroshima 739-8526, Japan}
\altaffiltext{3}{Department of Astronomy, Kyoto University,
Kitashirakawa-Oiwake-cho, Sakyo-ku, Kyoto 606-8502, Japan}
\altaffiltext{4}{National Astronomical Observatory of Japan, 2-21-1 Osawa, Mitaka, Tokyo 181-8588, Japan}
\altaffiltext{5}{Department of Physics, Faculty of Science and Engineering, Konan University, Kobe, Hyogo 658-8501, Japan}
\altaffiltext{6}{Institute of Astronomy, Graduate School of Science, The University of Tokyo, 2-21-1 Osawa, Mitaka,
Tokyo 181-0015, Japan}
\altaffiltext{7}{School of Science, Tokyo Institute of Technology, 2-12-1 Ohokayama, Meguro-ku, Tokyo 152-8551, Japan}
\altaffiltext{8}{Subaru Telescope, National Astronomical Observatory of Japan, 650 North A'ohoku Place, Hilo, HI 96720, USA}
\altaffiltext{9}{Graduate School of Science and Engineering, Saitama University,255 Shimo-Okubo, Sakura-ku, Saitama, 338-8570, Japan}
\altaffiltext{10}{Kavli Institute for the Physics and Mathematics of the Universe (WPI), The University of Tokyo Institutes for Advanced Study, The University of Tokyo, 5-1-5 Kashiwanoha, Kashiwa, Chiba 277-8583, Japan}
\altaffiltext{11}{Graduate School of Science and Engineering, Kagoshima University, 1-21-35 Korimoto, Kagoshima 890-0065, Japan
13 Gunma Astronomical Observatory, Takayama, Gunma 377-0702, Japan}
\altaffiltext{12}{Okayama Astrophysical Observatory, National Astronomical Observatory of Japan, National Institutes of Natural Sciences,
3037-5 Honjo, Kamogata, Asakuchi, Okayama 719-0232, Japan}
\altaffiltext{13}{Department of Astrophysics, Nagoya University, Furo-cho, Chikusa-ku, Nagoya, Aichi 464-8602, Japan}
\altaffiltext{14}{Koyama Astronomical Observatory, Kyoto Sangyo University, Motoyama, Kamigamo, Kita-Ku, Kyoto,
Kyoto, 603-8555, Japan}


\begin{abstract}
We present optical and near-infrared observations of a low-luminosity Type IIP supernova (SN) 2016bkv
from the initial rising phase to the plateau phase.
\blue{Our observations show that the end of the plateau is extended to
$\gtrsim \habs 140$~days since the explosion, 
indicating that this SN takes one of the longest time to finish the plateau phase.
among Type IIP SNe (SNe IIP), including low-luminosity (LL) SNe IIP.}
The line velocities of various ions at the middle of the plateau phase are as low as 1,000--1,500~\kms,
which is the lowest even among LL SNe IIP.
These measurements imply that the ejecta mass in SN~2016bkv
is larger than that of the well-studied LL IIP SN~2003Z. 
In the early phase, SN~2016bkv shows a strong bump in the light curve.
In addition, the optical spectra in this bump phase exhibit a blue continuum accompanied with a narrow \Ha emission line.
These features indicate an interaction between the SN ejecta
and the circumstellar matter (CSM) as in SNe IIn.
Assuming the ejecta-CSM interaction scenario,
the mass loss rate is estimated to be $\sim \habs 1.7 \times 10^{-2}~\Msun$~\years
within a few years before the SN explosion.
This is comparable to or even larger than the largest mass loss rate
observed for the Galactic red supergiants ($\sim \habs 10^{-3}~\Msun$~\years for VY~CMa).
We suggest that the progenitor star of SN~2016bkv experienced a
violent mass loss just before the SN explosion.
\end{abstract}


\section{Introduction}
Type IIP supernovae (SNe IIP) are characterized by hydrogen features
in their early-phase spectra and the `plateau' in their optical light curves
lasting for $\sim \hs 100$~days \citep{filippenko1997}.
SNe IIP represent the most common class of SNe and occupy 
$\sim \hs 70\% $ of core-collapse SNe within 60~Mpc
\citep[][]{li2011,graur2017,shivvers2017}.
The observational properties of SNe IIP show various diversities,
for example, in the durations of the plateau and the shapes of the initial light curves
\citep{anderson2014,faran2014,sanders2015,pejcha2015,valenti2016}.
So far, clear early-phase bump features have been found in several SNe II \citep[\eg][]{khazov2016}.
The early bump emission can be explained by the interaction between the SN ejecta 
and the circumstellar medium (CSM), which is originated from a stellar wind of the 
progenitor star \citep{moriya2017,morozova2017, yaron2017,dessart2017}.
These diversities may well be related to different natures of their progenitors.

The absolute magnitudes in the plateau phase also show a large variety,
ranging from $\sim \hs -14$ to $\sim \hs -17$~mag \citep{anderson2014}.
Note that this range is derived including some SNe IIL, which tend to show the brightest luminosities in the sample \citep[\ie $\sim \hs -17$~mag;][]{bose2015}.
In this paper, we refer to typical SNe IIP with the absolute plateau magnitudes of 
$\sim \hs -16$ or $-17$~mag as `normal SNe IIP'.
In the past $\sim \hs 20$~years, a subclass of faint SNe IIP has been recognized,
called low-luminosity SNe IIP \citep[LL SNe IIP;][]{pastorello2004}.
A typical luminosity of LL SNe IIP is lower than those of normal SNe IIP by 2--3~mag.
Typical velocities of the absorption lines are also lower than those of normal SNe IIP 
by a factor of 3--4 \citep{pastorello2004}.
The origin of these LL SNe IIP is still under debate:
They may be either originated from
weak explosions of the less massive stars ($\sim \hs 9~\Msun$) 
or failed explosions of the massive stars ($\sim \hs 25~\Msun$)
\citep{turatto1998, chugai2000, kitaura2006}.
To date, several LL SNe IIP have been well studied \citep[\eg ][]{turatto1998, pastorello2004, spiro2014, lisakov2017_08bk, lisakov2017_LL}.
However, the sample covering both the initial rising phase
and the tail after the plateau phase is still small.
Especially, the observations in near-infrared (NIR) wavelengths are still rare
for LL SNe IIP.

In this paper, we present our observations of a LL SN IIP~2016bkv.
SN~2016bkv was discovered by Koichi Itagaki on 2016 Mar 21.7 (UT) \citep{itagaki2016}.
At two days after the discovery, this SN was classified as an SN II,
by the spectral similarity to SN IIn 1998S \citep{hosse2016}.
Interestingly, this SN shows an unusually strong initial bump in the light curves.
The host galaxy, NGC~3184, is a face-on spiral galaxy and
known to be an active SN factory 
in the last 100~years (\eg SNe~1921B, 1921C, 1937F and 1999gi;
\citealt{leonard2002_99gi}),
The distance to the host galaxy is assumed to be $12.3\pm 2.2$~Mpc, 
adopting the mean redshift-independent distance reported in the literatures 
(via NED\footnote{http://ned.ipac.caltech.edu/}).
The explosion date is assumed to be 2016 Mar $20.4\pm 1.3$,
as constrained by the last non-detection date (Mar 19.1) by
Katzman Automatic Imaging Telescope (KAIT).
Throughout the paper, the observation epoch is given with respect to
the estimated explosion date.
\citet{hosse2018} presented optical light curves
and spectra of SN~2016bkv, and discussed the nature of the explosion
and progenitor using models of nebular spectra.
Our paper provides complementary data including optical and NIR light curves with a longer plateau coverage.

This paper is organized as follows.
In \S~2, we describe our observations and data reduction.
In \S~3 and 4, we show photometric and spectroscopic properties of SN~2016bkv, respectively.
We discuss the nature of this object in \S~5,
and finally, we give conclusions in \S~6

\section{Observations and Data Reduction}
\begin{figure}[h]
\centering
\includegraphics[width=8cm,clip]{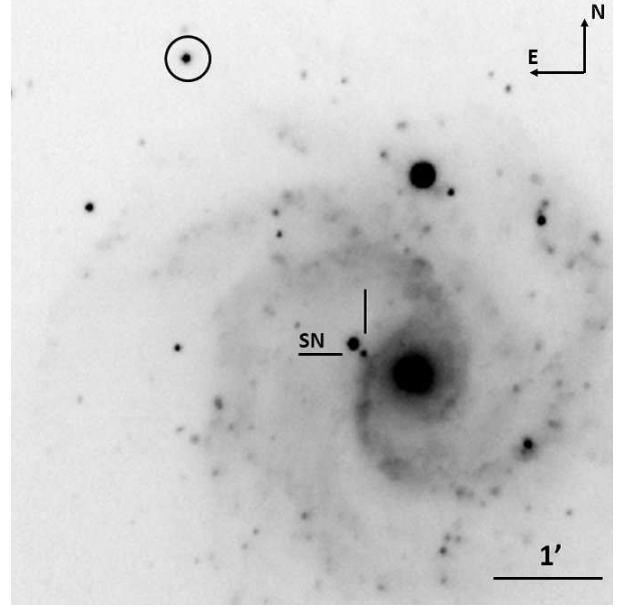}
\caption{$R$-band image of SN~2016bkv in NGC~3184 taken using HOWPol.
The location of SN and comparison star \citep{leonard2002_99gi} are marked.}
\label{fig:fc}
\end{figure}

\begin{figure}[h]
\centering
\includegraphics[width=8cm,clip]{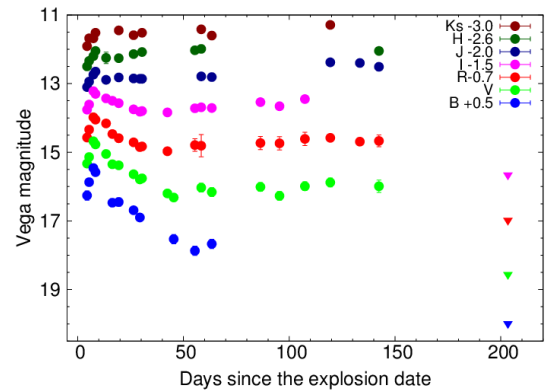}
\caption{Optical to NIR light curves of SN~2016bkv.
The Galactic interstellar extinction \citep{schlafly2011} has been corrected for.
The host-galactic extinction is estimated to be negligible (see \S 3)
and thus no correction has been made.}
\label{fig:lc}
\end{figure}

The optical imaging data were obtained using 
the Hiroshima One-shot Wide-field Polarimeter \citep[HOWPol;][]{kawabata2008}
and the Hiroshima Optical and Near-InfraRed Camera \citep[HONIR;][]{akitaya2014}.
These instruments were installed into the 1.5-m Kanata telescope at the 
Higashi-Hiroshima Observatory, Hiroshima University.
We obtained \bvri data using HOWPol in 10 nights from 2016 Apr 5.3 until Oct 9.4,
and also obtained \bvri data using HONIR in 14 nights from Mar 24.3 until Aug 9.1.
For the photometric measurements,
we adopted the Point Spread Function (PSF) photometry task in the {\it DAOPHOT} package 
\citep{stetson1987} equipped with the {\it IRAF} \citep{tody1986,tody1993}.
For the calibration of optical photometry, we used the magnitudes of a nearby comparison star given in \citet{leonard2002_99gi} (see Figure \ref{fig:fc}).
The derived optical magnitudes are summarized in Table \ref{table:opt}.
Figure \ref{fig:lc} shows the light curve in each band.
All magnitudes are given in the Vega magnitudes throughout the paper.

\begin{deluxetable*}{lllllll}
\tablewidth{0pt}
\tablecaption{Photometric observations of SN~2016bkv in optical bands.}
\tablehead{
  MJD &  Epoch     & B & V & R & I & Instrument\\
  &  (day)  & (mag) & (mag) & (mag) & (mag) & 
}
\startdata
57471.7 & 4.4 & 15.76(0.13) & 15.33(0.08) & 15.27(0.04) & 15.26(0.04) & HONIR\\
57472.7 & 5.4 & 15.37(0.07) & 15.14(0.06) & 15.04(0.04) & 15.11(0.04) & HONIR\\
57474.7 & 7.4 & 14.96(0.05) & 14.68(0.04) & 14.68(0.04) & 14.72(0.04) & HONIR\\
57475.7 & 8.4 & 15.08(0.04) & 14.77(0.05) & 14.75(0.04) & 14.80(0.05) & HONIR\\
57480.7 & 13.4 & --- & 15.05(0.05) & 14.86(0.04) & 14.93(0.08) & HONIR\\
57483.6 & 16.3 & 15.97(0.04) & 15.35(0.05) & 15.17(0.04) & 15.0(0.08) & HOWPol\\
57486.6 & 19.3 & 15.95(0.11) & 15.38(0.05) & 15.29(0.11) & 15.07(0.05) & HONIR\\
57493.6 & 26.3 & 16.19(0.04) & 15.64(0.05) & 15.41(0.05) & 15.25(0.05) & HONIR\\
57496.7 & 29.4 & 16.40(0.08) & 15.79(0.07) & 15.55(0.07) & 15.32(0.04) & HONIR\\
57497.7 & 30.4 & --- & 15.76(0.06) & 15.53(0.04) & 15.30(0.04) & HONIR\\
57509.7 & 42.4 & --- & 16.2(0.06) & 15.67(0.04) & 15.34(0.04) & HOWPol\\
57512.6 & 45.3 & 17.03(0.13) & 16.32(0.06) & --- & --- & HOWPol\\
57522.6 & 55.3 & 17.37(0.13) & --- & 15.49(0.18) & 15.22(0.06) & HONIR\\
57525.6 & 58.3 & --- & 16.03(0.12) & 15.51(0.32) & 15.19(0.05) & HONIR\\
57530.6 & 63.3 & 17.17(0.13) & 16.16(0.12) & --- & 15.21(0.08) & HONIR\\
57553.6 & 86.3 & --- & 16.01(0.12) & 15.43(0.19) & 15.04(0.08) & HOWPol\\
57562.5 & 95.2 & --- & 16.27(0.12) & 15.44(0.19) & 15.16(0.08) & HOWPol\\
57574.5 & 107.2 & --- & 15.99(0.12) & 15.31(0.19) & 14.95(0.08) & HOWPol\\
57586.5 & 119.2 & --- & 15.88(0.13) & 15.28(0.04) & --- & HONIR\\
57600.5 & 133.2 & --- & --- & 15.39(0.07) & --- & HONIR\\
57609.5 & 142.2 & --- & 15.99(0.18) & 15.37(0.17) & --- & HONIR\\
57670.8 & 203.5 & 19.49(0.13) & 18.56(0.18) & 17.68(0.17) & 17.16(0.08) & HOWPol\\
\enddata
\label{table:opt}
\end{deluxetable*}

In the late phase ($\gtrsim$150~days),
background emission from the host galaxy may contribute to the total measured flux.
To constrain possible contamination,
we use a pre-explosion image taken by the {\it Hubble Space Telescope (HST)} 
on 2001 Jan 24.
We found a probable stellar cluster at $m_{\rm F555W} = 19$--$20$ mag
within the typical PSF radius of our observations ($\sim \hs 2''$) around the SN.
Since the brightness in the plateau phase is much brighter than that of this source,
our magnitudes up to 150 days are reliable within a systematic error of
$\simeq$ 0.1~mag in optical bands.
On the other hand, the magnitudes in the last night (at 203~days)
are close to the background source, and thus we treat this brightness
as upper limits of SN fluxes.
Throughout the paper, we focus on the phases before 150 days
and our main discussion is not affected by the presence of 
the cluster-like component.

The \jhk imaging data were obtained using HONIR in 14 nights 
from Mar 24.3 until Aug 9.1.
We took the images using a dithering mode to accurately subtract the bright foreground sky.
We reduced the data according to the standard manner for the NIR data using
the PSF photometry method in {\it IRAF}.
The magnitude calibrations were performed using the magnitudes of nearby comparison stars
given in the 2MASS catalog \citep{persson1998}.
The derived \jhk magnitudes and their light curves are given in Table \ref{table:nir} and Figure \ref{fig:lc}, respectively.

\begin{deluxetable}{lllll}
\tablewidth{0pt}
\tablecaption{Photometric observations of SN~2016bkv in NIR bands.}
\tablehead{
  MJD &  Epoch     & J & H & Ks \\
  &  (day)     & (mag) & (mag) & (mag)
}
\startdata
57471.7 & 4.4 & 15.1(0.02) & 15.1(0.03) & 14.91(0.03)\\
57472.7 & 5.4 & 14.94(0.02) & 14.94(0.03) & 14.68(0.03)\\
57474.7 & 7.4 & 14.74(0.02) & 14.8(0.03) & 14.68(0.03)\\
57475.7 & 8.4 & 14.65(0.02) & 14.65(0.03) & 14.52(0.04)\\
57480.7 & 13.4 & 14.89(0.04) & 14.85(0.16) & ---\\
57486.6 & 19.3 & 14.82(0.02) & 14.86(0.05) & 14.45(0.07)\\
57493.6 & 26.3 & 14.85(0.03) & 14.74(0.03) & 14.59(0.04)\\
57496.7 & 29.4 & 14.86(0.02) & --- & ---\\
57497.7 & 30.4 & 14.86(0.02) & 14.68(0.03) & 14.52(0.04)\\
57522.6 & 55.3 & --- & 14.63(0.05) & ---\\
57525.6 & 58.3 & 14.79(0.05) & 14.59(0.02) & 14.42(0.06)\\
57530.6 & 63.3 & 14.81(0.04) & --- & 14.6(0.08)\\
57586.5 & 119.2 & 14.38(0.04) & --- & 14.29(0.08)\\
57600.5 & 133.2 & 14.4(0.04) & --- & ---\\
57609.5 & 142.2 & 14.51(0.03) & 14.65(0.06) & ---\\
\enddata
\label{table:nir}
\end{deluxetable}

\begin{figure}[h]
\centering
\includegraphics[width=8cm,clip]{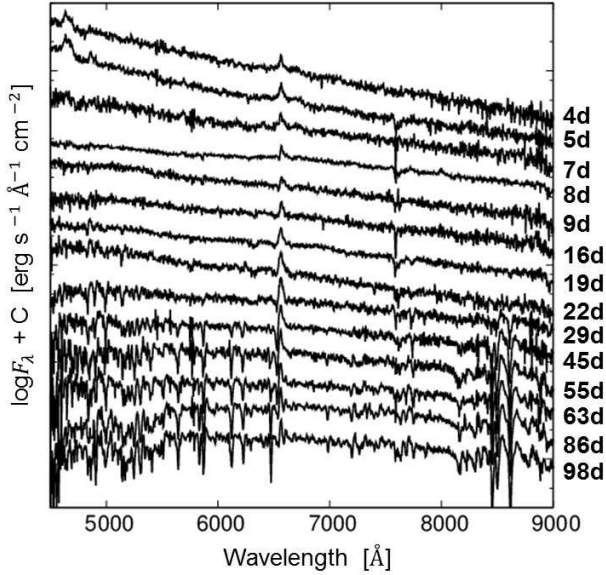}
\caption{Spectral evolution of SN~2016bkv. The epoch (days after the 
explosion; see \S~1) of each spectrum is given in the panel.}
\label{fig:spec}
\end{figure}

We performed optical spectroscopic observations using HOWPol
in 17 nights from Mar 24.3 until Jun 26.3.
We used a grism with a spectral resolution of $R\sim 400$
and a spectral coverage of 4500--9000~\AA.
We observed spectroscopic standard stars in the same nights for the flux calibrations.
For the wavelength calibration, we used the sky emission lines taken in the object frames.
The strong atmospheric absorption bands around 6900 \AA\ and 7600 \AA\ have
been removed using the smooth spectra of the hot standard stars.
The obtained spectra are shown in Figure \ref{fig:spec}, and the log of our spectroscopic 
observations is shown in Table \ref{table:spec}.

\begin{deluxetable}{lll} 
\tablewidth{0pt}
\tablecaption{Spectroscopic observations of SN~2016bkv.}
\tablehead{
  MJD  &  Epoch     & Exposure \\
  & (day) & (sec)
}
\startdata
57471.5 & 4.2 & 1600\\
57472.7 & 5.4 & 1400\\
57474.8 & 7.5 & 900\\
57475.6 & 8.3 & 1200\\
57476.7 & 9.4 & 1150\\
57477.7 & 10.4 & 1200\\
57483.6 & 16.3 & 1200\\
57486.5 & 19.2 & 1500\\
57489.5 & 22.2 & 1500\\
57493.5 & 26.2 & 1500\\
57495.6 & 28.3 & 1200\\
57496.6 & 29.3 & 2400\\
57512.7 & 45.4 & 1800\\
57522.6 & 55.3 & 1800\\
57530.6 & 63.3 & 2700\\
57553.5 & 86.2 & 2700\\
57565.5 & 98.2 & 2700
\enddata
\label{table:spec}
\end{deluxetable}

\section{Light curves}

\begin{figure}[h]
\centering
\includegraphics[width=8cm,clip]{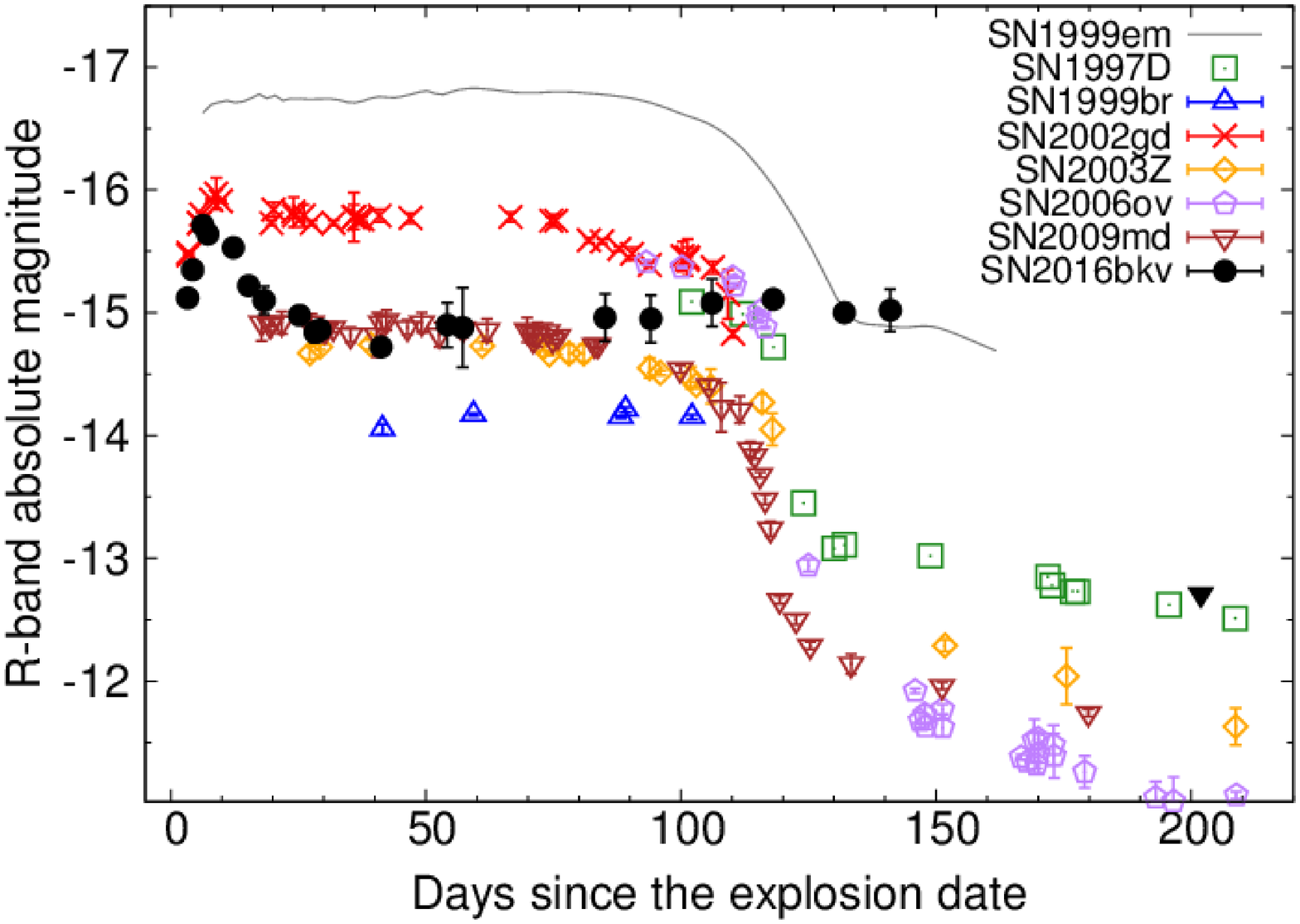}
\caption{The $R$-band light curve of SN 2016bkv compared with 
those of other LL IIP SNe~1997D, 1999br \citep{pastorello2004}, 
2002gd, 2003Z, 2006ov \citep{spiro2014}, 2009md \citep{fraser2011} and normal IIP SN~1999em \citep{leonard2002_99em}.
}
\label{fig:com_lc}
\end{figure}

The observed optical and NIR light curves of SN~2016bkv are shown in Figure \ref{fig:lc}.
The Galactic extinction of $E(B-V)=0.0144$ \citep{schlafly2011}
  has been corrected for.
We also derive an upper limit of the total extinction as $E(B-V) \lesssim 0.015$
by using the equivalent width of Na~\si~D absorption line \citep{poznanski2012}
in our stacked spectrum (EW $\lesssim 0.065$~\AA).
Since the Galactic extinction is close to the upper limit of the total
extinction, 
we consider that the interstellar extinction within the host galaxy is negligibly
small ($E(B-V)\lesssim 0.005$).

Figure \ref{fig:com_lc} shows the absolute $R$-band light curve of SN~2016bkv, 
compared with those of other LL SNe IIP and a normal SN IIP 1999em.
For the SNe other than SN 2016bkv,
we adopted the estimates of the explosion dates from \citet{pastorello2004},
\citet{fraser2011} and \citet{spiro2014} and
the total extinction corrections used in the references.
In the mid-plateau phase ($\sim \hs 80$~days),
the $R$-band absolute magnitude of SN~2016bkv is $M_R \sim \hs -15.0$~mag.
This is $\sim \hs 2$~mag fainter than that of the normal SN IIP 1999em.
Among LL SNe IIP,
SN 2016bkv is $\sim \hs 0.2$~mag brighter than SN~2003Z
and $\sim \hs 0.5$~mag fainter than that of SN~2002gd.
Note that, 
SN~2016bkv shows a slight brightening toward
$\sim \hs 120$~days, reaching $M_R\sim -15.1$~mag.

\blue{The light curves of SN~2016bkv have two remarkable characteristics.
One is that the end of the plateau phase ($\gtrsim \hs 140$~days after the explosion)
is later than those of other LL SNe IIP
for which good data covering the plateau available;
  $\sim \hs 110$~days in SNe 2003Z and 2009md \citep{spiro2014,fraser2011}.
In fact, this epoch is still later than that of SN~2009ib \citep{takats2015}, which shows the longest plateau length ($\sim \hs 130$~days) among normal SNe IIP.}

The other remarkable feature of SN 2016bkv is a bump in the early phase.
The light curves show a peak at $\sim \hs 7$~days after the explosion.
At 2--7~days, the SN shows a rapid brightening by about 0.8~mag, 0.65~mag, 
0.6~mag and 0.54~mag in the $B$-, $V$-, $R$- and $I$-band, respectively.
\blue{Such bumps in the light curves are also seen in the $V$-band light curve of SN 2002gd
\citep{spiro2014} and the $B$-band light curve of SN 2005cs \citep{pastorello2009} but the bump in SN 2016bkv is more pronounced.}
The bump gradually decays (by $\sim \hs 1$~mag through $\sim \hs 30$~days 
in the $R$-band) and then the SN seems to enter into the plateau phase.

\begin{figure}[h]
\centering
\includegraphics[width=8cm,clip]{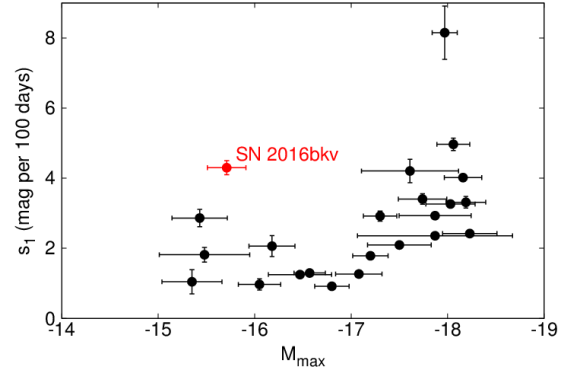}
\caption{Relation between absolute peak magnitudes in the $V$-band and
  the decay rates of the initial bumps (\sone) for 
  SN~2016bkv and other SNe II (except for type IIb and IIn).
  All the data except for SN~2016bkv are taken from \citet{anderson2014}.}
\label{fig:anderson}
\end{figure}

The decay rates of the bump \citep[\sone, as defined by][]{anderson2014} are
$\sim \hs 0.043$ and $\sim \hs 0.03$~mag day$^{-1}$ in the $V$-band and $R$-band, respectively.
The amplitudes of the bump and their decay rates are larger in shorter wavelengths
\citep{galbany2016}.
Figure \ref{fig:anderson} shows the peak absolute magnitude in the $V$-band ($M_{\rm max}$) 
and the decay rate of the bump (\sone) of SN~2016bkv as compared with those of other SNe II \citep{anderson2014}.
The error includes the uncertainty of the distance to the host galaxy 
and the uncertainty of a transition epoch from the bump into the plateau.
From Figure \ref{fig:anderson},
it is clear that
SN~2016bkv shows the fastest decay in SNe II having similar peak brightness or LL SNe IIP.
We will discuss the features and possible origin of the bump in \S 5.4.

\section{Spectra}

\begin{figure}[h]
\centering
\includegraphics[width=8cm]{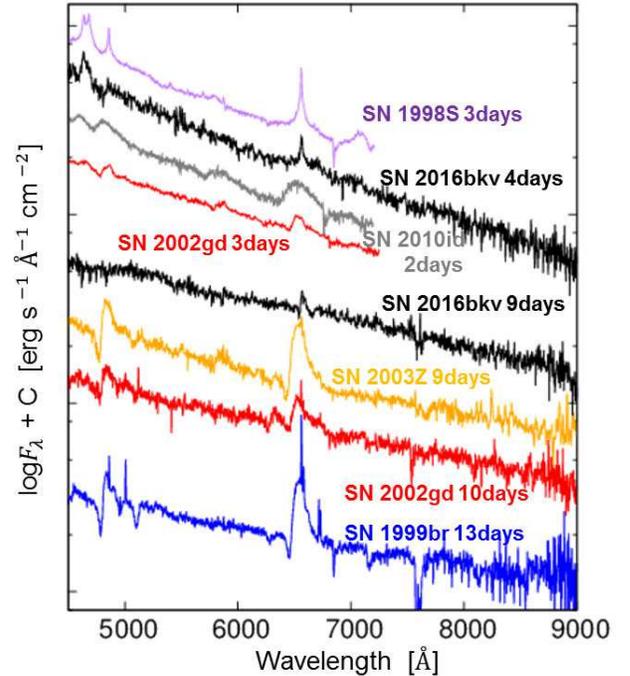}
\caption{Spectra of SN 2016bkv in the early phase in comparison with LL SNe IIP~2002gd, ~2003Z \citep{spiro2014}, ~1999br \citep{pastorello2004} and IIn 1998S \citep{fassia2001} at similar epochs. 
  The fluxes are shown in the logarithmic scale and arbitrarily scaled to avoid overlaps.}
\label{fig:com_early}
\end{figure}

\begin{figure}[h]
\centering
\includegraphics[width=8cm]{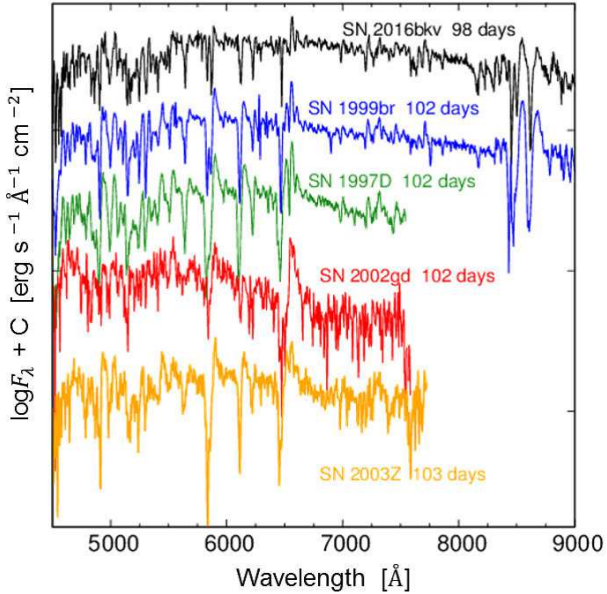}
\caption{Spectra of SN 2016bkv at 98 days in comparison with the spectra of LL SNe IIP~1999br, 
    ~1997D \citep{pastorello2004}, ~2002gd and ~2003Z \citep{spiro2014} at similar epochs
    (near the end of the plateau phase).}
\label{fig:com_sp}
\end{figure}

\subsection{Spectral Features}
\label{line_f}
Figure \ref{fig:spec} shows the spectral evolution of SN~2016bkv from 4 to 98~days.
Until 20~days, the spectra are dominated by a blue continuum,
superimposed by a weak and narrow ($\sim \hs 1000$~\kms) emission line of \Ha
(Figures \ref{fig:spec} and \ref{fig:com_early}).
\blue{After ~30 days after the explosion, various absorption lines,
\eg H, Sc~\sii, Fe~\sii, Ba~\sii, Ca~\sii, get stronger with time (Figure \ref{fig:spec}).}
No significant He~\si~ line is seen in the spectra.
In Figure \ref{fig:com_sp}, we compare the spectra of SN~2016bkv with those of LL SNe IIP at around 100~days.
At this epoch, the overall spectral features are similar to those in other LL SNe IIP, although SN~2016bkv shows slightly narrower features than the others.
At this epoch, SN~2016bkv shows a complicated structure of the \Ha line, which is also visible in other LL SNe IIP.
These facts confirm our classification of SN~2016bkv into LL SN IIP.

\begin{figure}[h]
\centering
\includegraphics[width=8cm,clip]{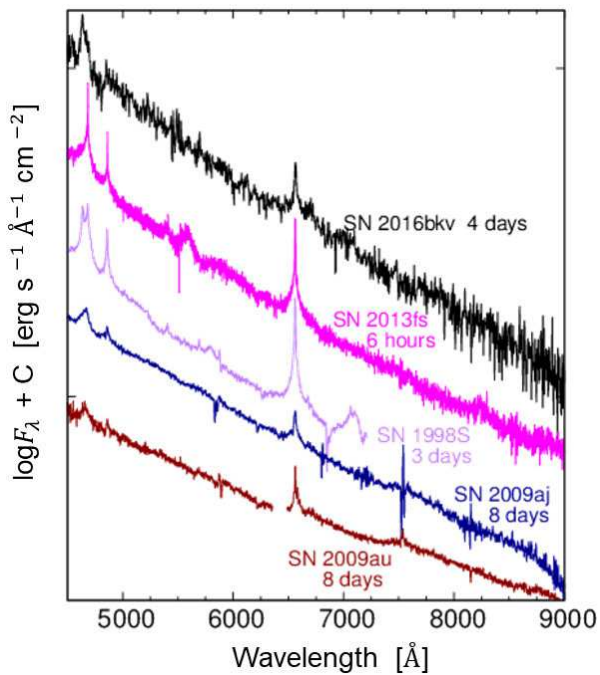}
\caption{Spectrum of SN~2016bkv at 4~days in comparisons with those of 
SNe 1998S \citep{fassia2001}, 2009aj, 2009au \citep{guti2017} and 2013fs 
\citep{yaron2017} at their earliest phases (0.3--8~days).}
\label{fig:fi}
\end{figure}

Although the overall spectral properties of SN 2016bkv show similarities
  to other LL SNe IIP, 
  \blue{SN~2016bkv shows some remarkable features in the earliest phases \citep[][,see Figure~\ref{fig:com_early}]{hosse2018}.}
The \Ha emission line at 4~days is narrower than in other LL SNe IIP,
with a possible asymmetric profile truncated at a redder side,
and there is no clear absorption component (see Figure \ref{fig:spec}).
At 8--9~days, a possible weak absorption component
in the bluer side of the \Ha is marginally visible 
\blue{(The detail will be described in \S 5.4).} 
After $\sim \hs 9$~days, the absorption component gradually develops
(Figure \ref{fig:spec}).
All of these properties are quite different from those in well-observed LL SNe IIP 2002gd,
 2003Z and 1999br  (see Figure \ref{fig:com_early}),
 which show clear P Cygni profiles in \Ha (and \Hb) at these epochs
 with larger line velocities.

Narrow Balmer emission lines
 with a blue continuum are sometimes seen in SNe II at their early phases.
Figure \ref{fig:fi} shows the spectral comparison of SN~2016bkv with other SNe II
with narrow emission lines and a blue continuum at 0--8~days.
In fact, the overall features of SN~2016bkv match well to those of
SN IIn 1998S \citep{fassia2001},
whose spectra show the emission lines of \Ha and \Hb, and
also the emission lines around 4600~\AA~
(a blend of highly-excitation lines of C~\siii, N~\siii~
and He~\sii~; \citealt{shivvers2017}).
As also shown in Figure \ref{fig:fi},
similar emission features are seen in SNe 2009aj,
2009au at comparable epochs \citep{guti2017}.
The similarity is also found for SN~2013fs \citep[iPTF13dqy;][]{yaron2017},
as the spectrum was observed much earlier since the explosion.
Given the epochs when these spectra are obtained,
the emissions from SNe 2009aj and 2009au may be powered by the CSM interactions,
while those of SN~2013fs are suggested to be flash ionized (FI) features \citep{galyam2014}.
We discuss the origin of these features from a viewpoint of 
possible CSM-ejecta interaction in \S \ref{emission} and \S \ref{absorption}.

\subsection{Line Velocities}
\label{line_v}
\begin{figure}[h]
\centering
\includegraphics[width=8cm,clip]{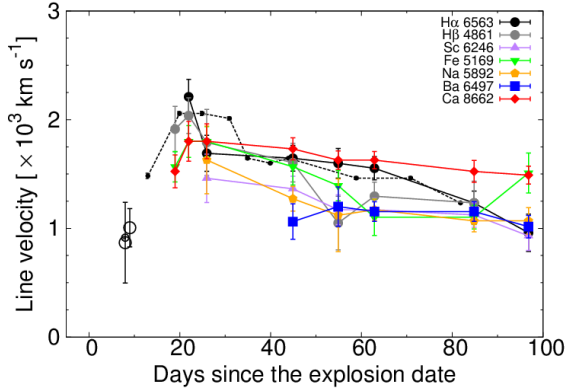}
\caption{Velocity evolutions of the \Ha, Sc~\sii, Fe~\sii, Na~\si~D, and Ca~\sii~ lines
  in SN~2016bkv.
The open black circles show the line velocity of the weak absorption component accompanied with the narrow \Ha~emission line seen at 8 and 9~days (see \S~4.2).
We obtain the errors by taking into account the uncertainties in the wavelength calibration and fitting error.
In addition to the measurements for our spectra, we also added
  \Ha line velocities measured in the spectra presented by Hosseinzadeh et al. (2018, dashed line).
}
\label{fig:line}
\end{figure}

After the end of the initial bump
($\sim \hs 30$~days after the explosion),
many resolved absorption lines are seen in the spectra (Figure \ref{fig:spec}).
We made line identifications by following \citet{pastorello2004}.
In LL SNe IIP, the \Ha absorption line around 50~days
is often affected by Ba~\sii~$\lambda$6497 because of the slight wavelength
difference between \Ha and Ba~\sii~$\lambda$6497 \citep{roy2011,lisakov2017_08bk}.
However, since the velocity of \Ha of SN~2016bkv is slower than that of other LL SNe IIP (see Figure \ref{fig:line}),
\blue{the profile of \Ha of SN~2016bkv can be clearly isolated from the Ba~\sii~$\lambda$6497
between 45 and 98~days.}
The line velocities were measured by fitting a Gaussian function to each 
absorption line.
Figure \ref{fig:line} shows the line velocity evolutions of 
various absorption lines: 
\Hb, Fe~\sii~$\lambda$5169,Na~\si~D, Sc~\sii~$\lambda$6246, 
Ba~\sii~$\lambda$6497, \Ha and Ca~\sii~$\lambda$8662. 
\blue{If we compare the lines created by the same element at the same ionization state,
the velocities and their evolutions are mutually consistent (\eg Ca~\sii~ NIR triplet).
Thus only representative line velocities are shown in Figure \ref{fig:line}.}
The velocities and their evolutions are consistent for each element;
the velocities are about 1,600--2,000~\kms at 22 days and 
evolves to 1,000--1,600~\kms at 85~days.
The Sc~\sii~ and Na~\si~D lines exhibit one of the lowest velocities among these lines.

In normal SNe IIP (\eg SN~1999em),
  the \Ha line velocities are always higher than the velocities of other elements
  at all the epochs \citep{leonard2002_99em}.
This trend is, however, not necessarily true for LL SNe IIP.
  For example, the faster line velocity is observed for the Ca~IR triplet and Fe~\sii~ than the \Ha in SN~2008in at 60-100 days after the explosion \citep{roy2011}.
Similarly, in SN~2016bkv,
  the Ca IR triplet line velocities are higher than the \Ha line velocity after $\sim \hs 80$~days.

\begin{figure}[h]
\centering
\includegraphics[width=8cm,clip]{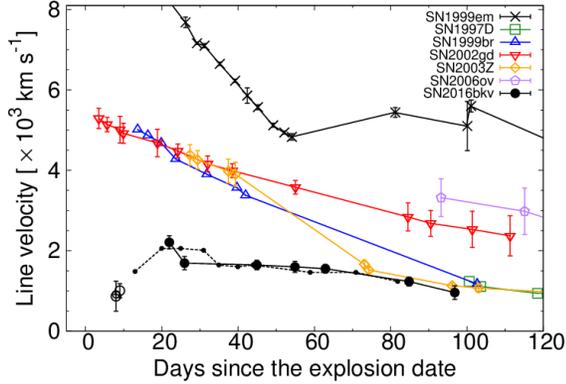}
\caption{Comparison of \Ha line velocities among LL IIP SNe~2016bkv, 1997D and 1999br \citep{pastorello2004}, ~2002gd, ~2003Z, ~2006ov \citep{spiro2014}, 
  and a normal SN IIP 1999em \citep{leonard2002_99em}.
The dashed line for SN 2016bkv shows the velocities measured
in the spectra by \citet{hosse2018}.}
\label{fig:com_ha}
\end{figure}

Figure \ref{fig:com_ha} shows the \Ha velocity evolution of SN~2016bkv,
compared with other LL SNe and the normal SN IIP~1999em \citep{leonard2002_99em}.
It is clear that LL SNe IIP commonly show lower velocities
than those of SN IIP 1999em over the entire phases.
SN~2016bkv exhibits the lowest velocities among LL SNe IIP until $\sim \hs 60$~days.
Interestingly, SN~2016bkv shows a slower evolution;
  the evolution of line velocities after $\sim \hs 20$~days is flatter than
  the other LL SNe IIP.
  The slow recession of the photospheric velocity
  means that the photospheric radius ($R_{\rm ph}$) tends to be large.
  This may be related to the slight increase in the luminosity at the end of
  the plateau phase ($L \propto R_{\rm ph}^2 T^4$) seen in SN~2016bkv.

The \Ha velocities at the earliest phases are of interest.
At 4, 5, and 7~days, the spectra exhibit the emission line of \Ha,
and then, a very weak absorption component is possibly detected at 8 and 9~days.
The velocities of these absorption components are measured as 800--1,000~\kms
(open symbols in Figures \ref{fig:line} and \ref{fig:com_ha}).
In these figures, we also added the measured velocities 
  using the spectra presented by Hosseinzadeh et al. (2018, dashed line).
In addition to the absorption lines,
  we also measure the width of the emission line of \Ha.
  We find that the FWHMs of the emissions are similar to
  the blueshift velocities of the absorption lines, i.e., 800--1,000~\kms.
These velocities are clearly slower than the velocities of the stronger absorption
component which appears after 26~days (1,700~\kms).
Implication from this feature is discussed in \S~\ref{absorption}.

\section{discussion}

\subsection{Bolometric Luminosity and $^{56}$Ni Mass}

\begin{figure}[h]
\centering
\includegraphics[width=8cm,clip]{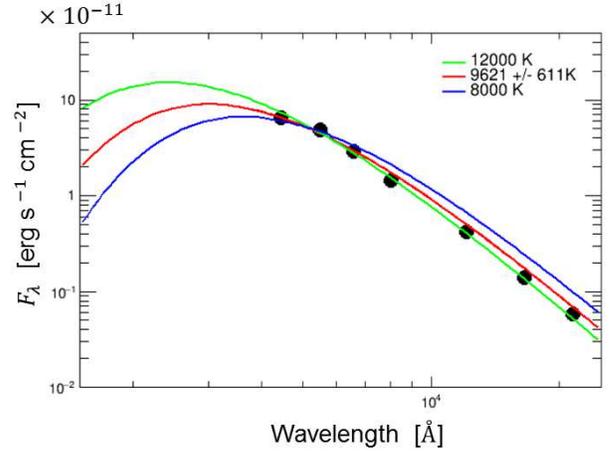}
\caption{SED of SN~2016bkv at 7.4~days after the explosion as compared to blackbody
  with different temperatures:
  the best fit model (in red), models of higher temperature (8000 K, in blue) and lower one (12000K, in green).}
\label{fig:bb}
\end{figure}

\begin{figure}[h]
\centering
\includegraphics[width=8cm,clip]{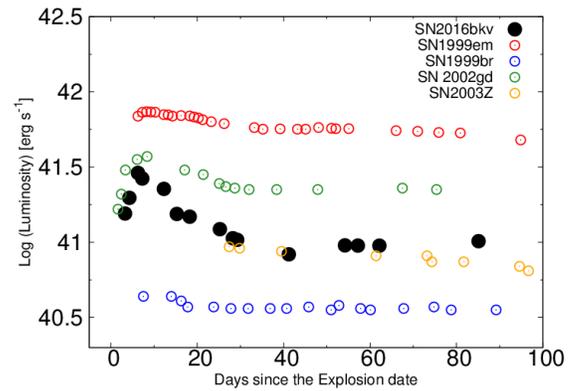}
\caption{Quasi-bolometric light curve of SN~2016bkv
  compared with that of LL SNe IIP 1999br \citep{pastorello2004}, 2002gd, 2003Z \citep{spiro2014} and a normal SN IIP 1999em \citep{leonard2002_99em}.
  For all the SNe, the quasi-bolometric luminosities are derived by
  blackbody fitting in the \bvri wavelength range.
  }
\label{fig:bol}
\end{figure}

We derived quasi-bolometric luminosities by integrating the \bvri fluxes as follows.
First, the extinction-corrected \bvri magnitudes were converted into fluxes
using filter parameters \citep{bessell1990} and the zero-magnitude fluxes
\citep{fukugita1995}.
Then, we fitted the spectral energy distributions (SEDs) with the blackbody function.
Finally the flux is integrated by using the best-fit blackbody parameters.
\blue{In the blackbody fit, we have included the \jhk if available,
in order  to constrain the temperature as accurately as possible.
We then integrated fluxes in the \bvri wavelength region.
The quasi-bolometric luminosities of other LL SNe IIP were also calculated in the same manner.
This allows fair comparisons to the other samples, as the NIR data are not always available.}

Figure \ref{fig:bb} shows the SED with $BVRIJHK_s$-band flux
and the best-fit blackbody function at 7.4~days after the explosion
(at the peak of the early bump).
The blackbody temperature is obtained to be $9,600 \pm 600$~K.
The temperature of LL SNe IIP 2002gd at 5.7 and 9.1~days are 12,700~K and 9,700~K, respectively \citep{spiro2014}.
Note that since there is no data at ultraviolet wavelengths
  and the temperature estimate in particular at $T > 10,000$ K is somewhat uncertain.
  To illustrate this uncertainty, the blackbody functions
  with the temperature of $\sim \hs 12,000$ and $\sim \hs 8,000$~K
  are also shown in Figure \ref{fig:bb}.
  This uncertainty affects the bolometric luminosities up to by a factor of about two.

Figure \ref{fig:bol} shows the quasi-bolometric light curve of SN~2016bkv,
compared with those of other LL SNe IIP and the normal SN IIP 1999em.
As in the multi-color light curves,
  the bolometric luminosity of SN~2016bkv clearly shows a bump ($\gtrsim$1~mag)
  in the early phase.
As expected, the luminosities of SN~2016bkv are always $\sim \hs 7-10$
times fainter than those of the normal SN IIP 1999em.
\blue{In comparisons with LL SNe IIP, 
the luminosities of SN~2016bkv are comparable to those of SN~2002gd before peak.}
After the bump ($\sim \hs 30$~days),
the luminosities of SN~2016bkv become 2--3 times fainter than those of SN~2002gd,
and become comparable to those of SN~2003Z.

Finally we give a constraint on the \Nifs mass of SN~2016bkv using the luminosity
at the tail.
As discussed in \S~2,
we only obtain an upper limit of the luminosity in the late phase
due to the background emission from the host galaxy.
By using the quasi-bolometric luminosity at 201~day
and assuming a relation between the tail luminosity
and the \Nifs mass suggested  by \citet{hamuy2003},
we obtain the upper limit of $M$(\Nifs)$~\lesssim 0.01\Msun$.
Our upper limit of the optical luminosity at $\sim \hs 200$~days
is close to the observed luminosity presented by \citet{hosse2018}.
Therefore, the ejected mass of \Nifs is, in fact, likely to be $M$(\Nifs)$~\sim \hs 0.01\Msun$. This is between the typical \Nifs masses of other LL SNe IIP and those of normal SNe IIP \citep{spiro2014}.

\subsection{Progenitor}
\label{progenitor}
The observed properties of SNe IIP reflect
the radius and mass of the progenitor and the energy of the explosion.
If the density structure and temperature are described in
a self-similar manner for different SNe,
the timescale of the plateau \blue{($t_{\EoP}$)}\blue{\footnote{In this context, we consider that the bump component is additional to the underlying plateau one. Thus, we define $t_{\EoP}$ as the epoch of the end of the plateau.} and the plateau luminosity ($L_{sn}$)}
are scaled as follows \citep{kasen2009}:

\begin{eqnarray}
t_{\EoP} \propto E^{-1/6} M_{\ej}^{1/2} R_0^{1/6} \kappa^{1/6} T_I^{-2/3} {\rm , and} \nonumber \\
L_{\rm SN} \propto E^{5/6} M_{\ej}^{-1/2} R_0^{2/3} \kappa^{-1/3} T_I^{4/3},
\label{eq:kasen}
\end{eqnarray}

\noindent
where $E$ is the kinetic energy of the ejecta,
$M_{\ej}$ is the ejecta mass, $R_0$ is the radius of progenitor star, 
$\kappa$ is the opacity, and $T_I$ is the ionization temperature.
Replacing the kinetic energy by the ejecta velocity ($v_{\ej}$),
using $E \propto M_{\ej} v_{\ej}^2$,
the relations (\ref{eq:kasen}) lead to the following expressions:

\begin{eqnarray}
t_{\EoP} \propto v_{\ej}^{-1/3} M_{\ej}^{1/3} R_0^{1/6} \kappa^{1/6} T_I^{-2/3} {\rm , and} \nonumber \\
L_{\rm SN} \propto v_{\ej}^{5/3} M_{\ej}^{1/3} R_0^{2/3} \kappa^{-1/3} T_I^{4/3}.
\label{eq:kasen2}
\end{eqnarray}

\noindent
Since it is a sound approximation that $\kappa$ and $T_I$ are the same for different SNe,
two quantities, $M_{\ej}$ and $R_0$, are constrained from
\blue{$t_{\EoP}$}, $L_{\rm SN}$, and $v_{\ej}$.

Here we estimate the properties of SN~2016bkv, using equations above.
We first use SN 2003Z for reference, whose properties have been extensively studied
with numerical light-curve models \citep[][$M_{\Z} = 11.3~\Msun, R_{\Z} = 260~\Rsun$]{pumo2017}.
As the photospheric velocity, we use the line velocity of the Sc~\sii~ line \citep{leonard2002_99em}.
The ratio of velocity of SN~2016bkv to that of LL SN IIP 2003Z
($v_{\bkv} / v_{\Z}$) is $\sim \hs 0.8$ around 60~days.
\blue{Using this velocity ratio, the ratio of $t_{\EoP}$ between SN~2016bkv ($t_{\bkv}$)
and SN~2003Z ($t_{\Z}$)
is expressed as follows:}
$t_{\bkv} / t_{\Z} \sim 1.1 (M_{\bkv} / M_{\Z})^{1/3} (R_{\bkv} / R_{\Z})^{1/6}$.
We can similarly derive the ratio of the luminosity as
$(L_{\bkv} / L_{\Z}) \sim 0.65 (M_{\bkv} / M_{\Z})^{1/3} (R_{\bkv} / R_{\Z})^{2/3}$.
From our data, $t_{\bkv} / t_{\Z}$ is given as $\sim$ 1.4 (see Figure \ref{fig:com_lc})
and $L_{\bkv} / L_{\Z}$ is $\sim$ 1.0 (see Figure \ref{fig:bol}).
Here we use $L_{\Z} = 1 \times 10^{41}$~\ergs and $t_{\Z} = 110$~days, respectively.
Therefore, we derive $M_{\bkv} / M_{\Z}\sim 1.7$ and $R_{\bkv} / R_{\Z}\sim 1.5$,
suggesting that the ejecta mass is larger in SN~2016bkv.

We also perform similar analysis using SNe 1999br and 2008bk
  \citep{pastorello2004,lisakov2017_08bk}.
  We obtain similar tendencies, \ie a larger ejecta mass than these LL SNe IIP,
  \ie 1.2 and 2.4 times larger than these objects, respectively.
  By adopting the explosion parameters presented by \citet{zampieri2003, lisakov2017_08bk, pumo2017},
  we derive the ejecta mass and progenitor radius of SN~2016bkv to be
  16.1--19.4~$\Msun$ and 180--1080~$\Rsun$, respectively \citep{spiro2014}.
  It should be noted, however, that since the methods to estimate
  the explosion parameters are different in different papers,
  the absolute values include large systematic uncertainties.

From these results, we suggest that
  the progenitor of SN~2016bkv is more massive than the progenitors of other LL SNe IIP.
It should be noted that \citet{hosse2018} suggest a relatively low
mass ($\sim \hs 9~\Msun$) progenitor from the nebular spectra,
in particular from the weakness of the [O~\si] line.
However, such a low-mass progenitor is unlikely to result
in a late $t_{\EoP}$ for $\gtrsim \hs 140$~days as derived for SN~2016bkv \citep[see][]{sukhbold2016}.
On the other hand, our estimate using the properties observed of the plateau phase
has a different possible systematic error,
based on the assumption that the same scaling relations of normal SNe IIP
(Eqs. \ref{eq:kasen} and \ref{eq:kasen2}) are applicable to LL SNe IIP.
Therefore, the derived properties are subject to uncertainties.
We need more detailed, consistent modeling of both light curves and spectra
from the early to nebular phases.

\subsection{Properties of CSM Inferred from the Early Bump Emission}
\label{emission}
SN 2016bkv shows a clear bump in the optical light curves in the early phases.
In addition, the early spectra show a narrow \Ha emission line with a blue continuum.
These properties suggest the interaction
between the SN ejecta and the circumsetellar matter (CSM)
as in SNe IIn.
The signatures of the interaction are seen until $\sim \hs 20$~days.

The luminosity arising from the interaction between the SN ejecta and CSM
can be estimated as
\begin{equation}
L= 4\pi r_{s}^2 \cdot \frac{1}{2}\rho v_{\ej}^3 \epsilon,
\end{equation}
where $r_{s}$ is the radius of the shock and $\epsilon \sim 0.1$ is a
conversion factor from the kinetic energy to radiation \citep{moriya2014}.
At the peak of the bump, 7.4 days after the explosion,
the luminosity of the bump component is $L = 4 \times 10^{41}$~\ergs.
Here we extract the luminosity of the bump component
from the total luminosity at 7--9~days
by subtracting the plateau luminosity at 40--60~days
(assuming that the SN luminosity is constant).
The plateau component accounts for about one third of the total flux
(see Figure \ref{fig:bol}).
Assuming that the ejecta velocity is $v_{\ej}\sim \hs 2,000$~\kms,
  which is observed in \Ha absorption at 22~days (Figure \ref{fig:line}),
  the radius reaches $r_{s} \sim 1,800~\Rsun$ at the peak of the bump.
Then we obtain the mass loss rate of $\dot{M}_{0} / u_{v, 5} = 1.7 \times 10^{-3}$~$\Msun$ \years where $\dot{M}_{0}$ is the mass loss rate and $u_{v,5}$ is the wind velocity normalized to $10^5$\cms.
The CSM mass within $r_{s}$ is $M_{\csm} \sim 6.8 \times 10^{-2}~\Msun$.

If we assume a typical velocity of the red supergiant (RSG) wind,
$u_{v, 5} = 10$ (\ie 10~\kms),
the mass loss rate is $\dot{M} \sim 1.7 \times 10^{-2}~\Msun$~\years.
Then, the CSM within the shock radius $r_{s}$ corresponds
  to the mass loss in the final 4.1 years.
These results are consistent with those presented
by other studies for the SNe II showing early bumps
\citep[\eg][]{morozova2017,moriya2017},
suggesting that the progenitor of SN~2016bkv has also experienced
an intensive mass loss just before the SN explosion.

The dense CSM derived above might also follow a strong flash ionized (FI) features
\citep[\eg][]{galyam2014,khazov2016},
if there would have been a sufficiently strong ionizing flux
\footnote{We use the terminology of FI features
    for the features produced by radiative ionization from
    shock breakout either at the stellar surface or at the dense CSM.}.
However, at the peak of the bump, this effect is likely insignificant already,
since the FI features are typically found only within the first day
as seen in SN~2013cu and some other SNe \citep{galyam2014}
and soon disappear following the decay of the UV flux from an SN.
In addition, SN~2016bkv is a LL SN IIP and the ionizing photon
would be less intense than those in normal SNe IIP, leading to fainter FI features.
Therefore, we do not attribute the emission lines in the early phase
to the FI features.
It is possible that SN~2016bkv might have shown a genuine FI feature,
if it would have been observed within the first day.
Adopting the properties of CSM as derived above,
we estimate that a strong \Ha emission
($L_{H\alpha} \sim 7\times10^{40}$~\ergs) should have existed
if such an observation would have been performed within a day after the explosion.
Interestingly, SN~2013cu and SN 2016bkv show common emission lines
of C \siii/N \siii~ at $\sim \hs 3-4$~days. These features may be caused by
ionizing photons from the on-going interacting region.

\subsection{Possible Eruptive Mass Loss}
\label{absorption}
During several days just after the bump peak (7--10~days after the explosion),
spectra of SN~2016bkv show a hint of a weak \Ha absorption with the velocities of 800--1000~\kms.
The origin of this feature could be distinct from the strong P Cygni features with the 
broader emission lines after $\sim \hs 26$~days (Figure \ref{fig:spec}),
which are produced by the SN ejecta.
As discussed in \S~\ref{emission}, the bump in the light curve and narrow emission lines
at the early phases are likely to be caused by the CSM interaction.
In this context, we speculate that the absorption component might be
originated in the unshocked CSM.
The observed velocities (800--1,000~\kms) are much higher than a typical wind velocity of RSGs.
This suggests that the progenitor star might have experienced a sudden,
eruptive mass loss ejecting $\sim 6.8 \times 10^{-2}~\Msun$ within $r_{s}$
at only 0.2~years prior to the explosion.
Although it is rather speculative, it may be the case
given the large uncertainty in the mass loss of the progenitor just before the explosion.
However, the gradual increase of the \Ha velocities from 1,000 to 2000 \kms (Figure \ref{fig:line}),
is not well explained if the low velocity component is attributed
to unshocked CSM.
Dense observations at the early phases are necessary to
fully understand possible activities of the progenitor of LL SNe IIP.

\section{Conclusions}
We present our optical and NIR observations of the low-luminosity Type IIP SN~2016bkv.
Our observations cover the initial rising phase through the plateau phase,
and make this object one of the best observed LL SNe IIP.
\blue{The end of the plateau is the latest ($\gtrsim \hs 140$~days) among other LL SNe IIP.}
In the initial phase, SN~2016bkv clearly shows a bump ($\gtrsim$1~mag) in the light curves.
The bump peaks at $\sim \hs 7$~days after the explosion.
During this initial phase, optical spectra show a blue continuum
and narrow H emission lines.
These features suggest the interaction between the SN ejecta and the CSM.

\blue{Using the scaling relations for the timescale of the plateau and its luminosity,}
the ejecta mass of SN~2016bkv is estimated to be larger than those
of other LL SNe IIP.
From the luminosity of the early bump and the \Ha line emission,
we conclude that the progenitor star has a relatively massive CSM
in the vicinity of the progenitor star.
If the CSM has been produced by the stellar wind in the progenitor,
the required mass loss rate is estimated to be
$\sim \hs 1.7 \times 10^{-2}~\Msun$~\years
and the accumulated mass of the CSM is
$\sim \hs 6.8 \times 10^{-2}~\Msun$ at least within
several years before the SN explosion (for the assumed wind velocity of 10~\kms).
The mass loss rate inferred for SN~2016bkv is comparable to or even larger than
the largest mass loss rate estimated for RSG stars in the Milky Way
(\eg $\sim \hs 10^{-3}~\Msun$~\years for VY CMa; \citealt{smith2009}).

\acknowledgments
This research has made use of the NASA/IPAC Extragalactic Database (NED) which is operated by the Jet Propulsion Laboratory, California Institute of Technology, under contract with the National Aeronautics and Space Administration. 

This research has made use of the NASA/ IPAC Infrared Science Archive, which is operated by the Jet Propulsion Laboratory, California Institute of Technology, under contract with the National Aeronautics and Space Administration.

KM acknowledges support provided by Japan Society for the Promotion of Science (JSPS) through KAKENHI Grant 17H02864.


\end{document}